\documentstyle[twocolumn,psfig]{MN}
\title[Spectra of cores and hotspots] {Relativistic beaming effects in 
the spectra of cores and hotspots in radio galaxies and quasars}

\author[C.H. Ishwara-Chandra and  D.J. Saikia]
{C.H. Ishwara-Chandra$^{1\,,2\,,3}$\thanks{E-mail: ishwar@ncra.tifr.res.in} and 
D.J. Saikia$^{1}$\thanks{E-mail: djs@ncra.tifr.res.in}\\
$^1$ National Centre for Radio Astrophysics, TIFR, Post Bag 3, 
Ganeshkhind, Pune 411 007, India \\
$^2$ Joint Astronomy Programme, Department of Physics, Indian 
Institute of Science, Bangalore, 560 012, India \\
$^3$ Raman Research Institute, Sadashiva Nagar, Bangalore 560 080, India
}
 
\date{}
\begin{document}
\maketitle
\begin{abstract}
We present the results of multi-wavelength observations of cores
and hotspots at L, C, X and U bands 
with the Very Large Array, of a matched sample of
radio galaxies and quasars selected from the Molonglo Reference 
Catalogue. We use these observations to determine the spectra of 
cores and hotspots, and test the unified scheme for radio galaxies
and quasars. Radio cores have been 
detected at all wavelengths in all the quasars in our sample, 
while only $\sim$ 50 \% of the galaxies have detected cores in 
at least one wavelength. The degree of core prominence in this 
sample is consistent with the unified scheme for radio galaxies 
and quasars. A comparison of the distributions of the two-point 
spectral index of the cores in our sample of lobe-dominated quasars 
(LDQs) with a matched sample of core-dominated quasars (CDQs) shows 
that the distributions for these two classes are significantly different,
consistent with the expectations of the unified scheme.
The difference in the spectral indices of the two hotspots on opposite sides
are also significantly larger for quasars than radio galaxies,
as expected in the unified scheme. We also investigate the  
relationship between the spectral index of the hotspots and redshift or 
luminosity for our sample of sources.

\end{abstract}


\section{Introduction}
In the orientation-based unified scheme for radio galaxies and
quasars, these objects are believed to be intrinsically similar,
but appear to be different because of their different orientations to the
line-of-sight. The quasars are oriented close to the line-of-sight,
while the radio galaxies lie close to the plane of the sky, the
dividing line being about 45$^\circ$. There has been a 
reasonable degree of observational evidence in different wave bands
in support of such a scheme (Scheuer 1987; Barthel 1989; 
Antonucci 1993; Urry \& Padovani 1995).

In this paper we investigate the effects of
relativistic beaming in the spectra of cores and hotspots in a
matched sample of radio galaxies and quasars, to further test the
unified scheme for these objects. We have attempted
to determine the spectra of the cores using observations at 
two widely-spaced L-band frequencies, and at C, X and U bands. 
These observations  represent one of the more extensive studies 
on the spectra of relatively weak cores in lobe-dominated radio 
sources. In an earlier study, Athreya et al. (1997) found a high 
proportion of cores in their sample of high-redshift radio 
galaxies to have a steep radio spectrum between 4.7 and 8.3 GHz,
and identified this with the optically thin part of the core spectrum.
They also examined the spectra of core- and lobe-dominated quasars
from available data in the literature, and suggested that the 
galaxy and quasar cores have intrinsically similar radio spectra. 
Lonsdale, Barthel \& Miley (1993) found most of the radio cores 
in their sample of high-redshift, lobe-dominated quasars to have
steep spectra between 5 and 15 GHz, indicating the presence of 
optically thin components. This is possibly due to the high
emitted frequencies corresponding to these sources.
We present the spectra of the cores in our sample 
and examine the effects of relativistic beaming in Section 4.

Relativistic effects in the spectra of hotspots in radio galaxies and
quasars selected from the 3CR sample have been examined recently
by Dennett-Thorpe et al. (1997, 1999). For the quasars they find that
the spectrum of the high surface brightness region is indeed flatter
on the jet side, but the spectrum of the low-brightness region is
flatter on the side with the longer lobe. They interpret these results
in terms of relativistic bulk motion in the spectra of the high
brightness regions, and differential synchrotron ageing in the low-brightness
regions. For the FRII galaxies, the spectral asymmetries appear to
be uncorrelated with jet sidedness at all brightness levels, but appear
to be related to relative lobe volume. In addition, earlier studies have
shown that the less depolarized lobe has a flatter radio spectrum
(Liu \& Pooley 1991), and generally faces the radio jet (Laing 1988;
Garrington et al. 1988; Garrington, Conway \& Leahy 1991). 
We examine the effects of relativistic beaming in the spectra of the 
hotspots in our sample of radio galaxies and quasars using the scaled-array
observations at L, C and U bands. These results are presented in Section 5.
We also investigate any dependence of hotspot spectral indices on redshift or 
radio power. 

In this paper, the sample and the observations are described in Section 2,
while the observational results on the radio cores are presented in Section 3.
The observed core prominence, the core spectra and their consistency with the
unified scheme are presented in Section 4. The consistency of the hotspot
spectra with the relativistic beaming effects and the unified scheme, as
well as the dependence of hotspot spectra on redshift or luminosity are
discussed in Section 5. The conclusions are summarised in Section 6.

\begin{table}
\caption{ Observing log }
\begin{tabular}{l l l l l }

Array  & Obs.  &Obs.     & Band-  & Date of obs.  \\
Conf.  & band  &Freq.    & width  &       \\
       &       & MHz     & MHz    &        \\
       &       &         &        &         \\
BnA    & L     &1365   &  50 &  1995 Sep 20  \\
       & L     &1665   &  25 &                \\
CnB    & C     &4635   &  50 &  1996 Jan 20,31 \\
       & C     &4935   &  50 &                  \\        
DnC    & U     &14965  &  50 &  1997 Sep 15, 16  \\
DnC    & U     &14965  &  50 &  1997 Oct 3, 4, 12 \\
BnA    & X     &8447   &  25 &  1997 Feb 3  \\
\end{tabular}
\end{table}

\section{Source Sample and observations}
The sample and the selection criteria are discussed in detail 
in an earlier paper reporting the polarization properties of this
well-defined sample of radio galaxies and quasars selected
from the Molonglo MRC/1Jy sample
(cf. Ishwara-Chandra et. al 1998, hereinafter referred to as IC98). 
Briefly, the sample consists 
of 15 quasars and 27 radio galaxies larger than about an 
arcminute in size with similar redshift, luminosity and projected linear size 
distributions. The observations 
were made with scaled arrays of the Very Large Array (VLA) of the 
National Radio Astronomy Observatory at 1.4 and 1.7 GHz (L-band), 
5 GHz (C-band) and 15 GHz (U-band) 
with resolutions of $\sim$5$^{\prime\prime}$, while observations at 8 GHz (X-band)
have a resolution of $\sim$1$^{\prime\prime}$. At 15 GHz, 14 quasars and 
10 radio galaxies were observed, while at 8 GHz only those sources in the RA 
range 03h to 13h were observed due to scheduling constraints. The observing log is 
summarized in Table 1. All the data were calibrated and analyzed 
using the NRAO {\tt AIPS} package. The images at X band and U band 
were corrected for primary beam attenuation. The U-band images have been
restored with the same resolutions as the L- and C-band images, which are
listed in Table 3 of IC98.

%

\begin{table*}
\caption{Observed and derived parameters of the radio cores}
\begin{tabular}{ l l l r r r r r r r r l l }
Source     &Id &  z    & \multicolumn{6}{c}{Flux density of the core } & $\alpha_L^C$ & $\alpha_C^U$ & f$_c$ & Notes \\
Name       &   &       & 1365 & 1665  & 4635 & 4935 & 8450 & 15000 &       &       &      &  \\
	   &   &       & MHz  & MHz   & MHz  & MHz  & MHz  &  MHz  &       &       &      &  \\
           &   &       &      &       &      &      &      &       &       &       &      &  \\
0017$-$207 & Q & 0.545 & 2.8$^d$  &  2.8  & 2.2  & 2.1  &      & 1.2   & 0.25  & 0.51  & 0.017  &   \\ 
0058$-$229 & Q & 0.706 & 4.3$^d$  &  3.8$^d$  & 2.1  & 2.4  &      & 1.0   & 0.51  & 0.71  & 0.020  & \\ 
0133$-$266 & Q & 1.53  & 14.1 &  13.2 & 16.6 & 16.8 &      &       &$-$0.17&       & 0.10  & \\
0148$-$297 & G & 0.41  & 1.9  &       & 0.4  &      &      & 2.5   &       &       & 0.00060   &  N \\
0428$-$281 & G &0.65 &      &       &$\sim$0.3   & $\sim$0.2  &  $\sim$0.3 &       &  &  & 0.00074   & N \\
           &   &       &      &       &      &      &      &       &       &       &         &  \\
0437$-$244 & Q & 0.84  & 16.9 &  15.2 & 12.8 & 12.5 & 10.5 & 4.4   & 0.21  & 0.92  & 0.089   &  \\
0454$-$220 & Q & 0.533 & 190.7&  180.5& 137.0& 137.5& 167.4& 162.1 & 0.26  &$-$0.15& 0.26  &  \\
0551$-$226 & G & 0.8 &      &       &      &      & 0.4  &       &       &       & 0.0040   &  N \\
0938$-$205 & G & 0.371 &      &       &      &      & 0.6  &       &       &       & 0.0058   &    \\
0955$-$283 & G & 0.8 &      &       &      & 0.6  & 1.0  &       &       &       & 0.0042   &  \\
           &   &       &      &       &      &      &      &       &       &       &         &  \\
1022$-$250 & G & 0.34&      &       &      &      & $\sim$0.3  &     &       &       & 0.0039   &  N core?  \\
1023$-$226 & G &0.586&      &       &      &      & 0.4  &       &       &       & 0.0050   &    \\
1025$-$229 & Q & 0.309 & 11.9 & 10.0  & 10.2 &10.2  & 9.9  & 7.9   & 0.07  & 0.22  & 0.09   &   \\
1026$-$202 & G & 0.566 &      &       & 0.7  &  0.7 & 0.6  & 0.7   &       &       & 0.0040   &  \\
1029$-$233 & G &0.611&      &   0.8 & 0.5  &  0.4 & 0.5  &       & 0.56  &       & 0.0036   &  \\
           &   &       &      &       &      &      &      &       &       &       &         &  \\
1052$-$272 & Q & 1.103 & 2.3  &  2.1  & 1.8  & 2.0  & 1.2  & 1.2   & 0.13  & 0.40  & 0.0098 &  \\
1107$-$218 & G &1.5  &      &       &      &      & $\sim$0.3  &     &       &       & 0.0023   &  N core?  \\
1126$-$290 & G & 0.41&      &       & 1.9  & 1.7  & 1.4  &       &       &       & 0.0054   &  N \\
1226$-$297 & Q & 0.749 & 1.7  &  2.3  & 4.0  & 3.9  & 4.6  & 4.6   &$-$0.61&$-$0.13& 0.029  &  \\
1232$-$249 & Q & 0.352 & 17.5 &  14.7 & 11.2 & 10.6 & 10.0 & 8.2   & 0.34  & 0.25  & 0.023  &  \\
           &   &       &      &       &      &      &      &       &       &       &         &  \\
1247$-$290 & Q & 0.77  & 14.2$^d$ &  11.4$^d$ & 2.4  & 2.1  & 1.7  & 1.2   & 1.50  & 0.55  & 0.013  &  \\
1257$-$230 & Q & 1.109 & 16.2 &  16.4 & 15.4 & 15.0 & 10.1 & 8.1   & 0.06  & 0.55  & 0.055  &  \\
1358$-$214 & G & 0.5   &      &       &      &      & 0.6  &       &       &       &  0.0074  &    \\
2035$-$203 & Q & 0.516 & 34.7 &  38.1 & 40.0 & 38.7 &      & 29.7  &$-$0.073& 0.25 & 0.20 &  \\
2040$-$236 & Q & 0.704 & 119.1&  116.4& 91.7 & 90.3 &      & 74.3  & 0.22  & 0.18  & 0.740  &  \\
           &   &       &      &       &      &      &      &       &       &       &         &  \\
2118$-$266 & G & 0.343 & 27.2 &  28.6 & 51.6 & 51.8 &      & 47.5  &$-$0.53& 0.074 &  0.620  &  \\
2213$-$283 & Q & 0.946 & 38.3 &  40.7 & 54.2 & 54.1 &      & 56.7  &$-$0.27&$-$0.04& 0.19  &  \\
2311$-$222 & G & 0.434 & 2.4  &  1.6  & 1.9  & 1.7  &      & 1.1   & 0.11  & 0.43  &  0.0077  &  \\
2338$-$290 & Q & 0.446 & 30.4 &  31.9 & 40.4 & 40.8 &      & 64.5  &$-$0.23&$-$0.40& 0.40  &  \\
\end{tabular}
\end{table*}

\section{Detection of radio cores}
All the quasars have detected radio cores at all wavelengths. 
In the case of radio galaxies, only two have reliable cores at 
all wavelengths, but many of them have weak cores in at least 
one wavelength. Of the 27 radio galaxies, 14 of them have a core 
detected in at least one wavelength, only 5 of which have a 
core flux density greater than 1 mJy. These 14 include two
possible cores which are weak features seen in only the X-band images 
close to the positions of the optical galaxies. The peak flux 
densities of cores have been used to minimise contribution from nearby
extended emission, and these values have been estimated using the {\tt AIPS} 
task {\tt IMFIT}. In the giant source 1025$-$229, the  
core flux densities have been estimated by fitting a gaussian to
both the core and the nearby jet-like component in the L-, C- and
U-band images. All the core flux densities estimated from the Gaussian fits
are similar to the values at the pixel of maximum brightness near
the optical position. 
Only those cores which could be clearly identified in the images have been
listed in Table 2, which is arranged as follows. 
Column 1: Source name; column 2: optical identification where Q denotes a
quasar and G a radio galaxy; column 3: redshift of the source; 
columns 4 to 9: peak flux densities of the cores at 
1365, 1665, 4635, 4935, 8450 and 15000 MHz in units of mJy. In a 
few cases the core flux density could be significantly over-estimated
due to extended diffuse emission in its vicinity. These have been
marked with a superscript $d$. 
Columns 10 and 11: spectral indices between L and C bands, and between
C and U bands, where the spectral index, ${\alpha}$, is defined as 
S$(\nu)\propto\nu^{-\alpha}$.
The spectral indices have been estimated using linear least square fits
between the bands using the core flux densities at both the frequencies in
the L and C bands. Cores with the superscript $d$ have not been used.
Column 12: the fraction of emission from the core at an emitted frequency
of 8 GHz using the core spectral index determined from our observations
and a spectral index of 1 for the extended emission.
Column 13: N indicates that there is a note on the source in the text,
while core? indicates a possible core. The positions of the optical objects
have been listed by either McCarthy et al. (1996, hereinafter referred to
as M96) if it is a galaxy, or by Kapahi et al. (1998, hereinafter referred
to as K98) if it is a quasar. The typical errors in the optical positions
are about 1.$^{\prime\prime}$5  (M96) while for the quasars it is about 0.$^{\prime\prime}$5
(K98). The positions of the new radio cores, which are not listed in K98,
are presented in Table 3.

\subsection*{Notes on individual sources:}

\noindent {\it 0148$-$297:} The core is relatively weak at 4635 MHz;
its position is the same as that of the stronger core at 15 GHz, and also
consistent with the one seen at 1365 MHz. 

\noindent {\it 0428$-$281:} A weak radio core is detected at both the
C bands as well as the X band. The radio positions of the core are the same 
in all the frequencies; but the core is about 5.$^{\prime\prime}$9 from the optical 
position given by M96. 

\noindent {\it 0551$-$226:} The radio core detected at X band is about
2.$^{\prime\prime}$9 from the optical position listed by M96. 

\noindent {\it 1022$-$250:} The possible radio core is about 2.$^{\prime\prime}$1 
from the position of the optical galaxy given by M96.

\noindent {\it 1107$-$218:} The possible radio core is about 1.$^{\prime\prime}$6
from the position of the optical galaxy given by M96.

\noindent {\it 1126$-$290:} The positions of the core at both the C bands and
at the X band are consistent, but the core is about 12.$^{\prime\prime}$4 
from the optical position given by M96. The position of the centroid of the optical
galaxy measured from recent CCD observations of the field is 
RA: 11$^h$ 26$^m$ 26.$^s$434 and Dec: $-$29$^\circ$ 05$^\prime$ 02.$^{\prime\prime}$74
in B1950 co-ordinates. This is close to the position of the radio core.

\begin{figure}
\vbox{
\vspace{-1.0in}
\psfig{figure=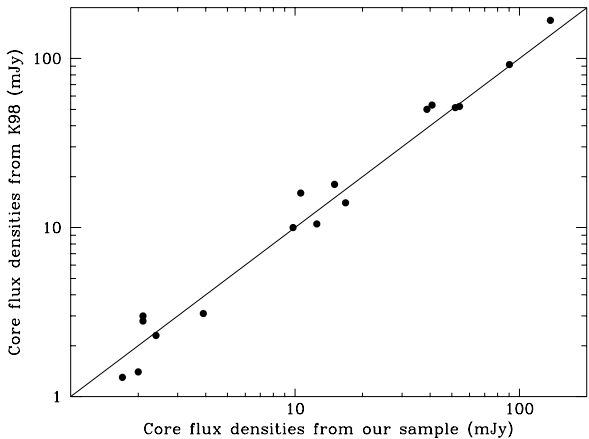,height=3.25in}
\vspace{0.2 in}
}
\caption{Plot of core flux densities at 4.9 GHz from K98 against 
the values estimated from our observations.  Our estimates of core 
flux densities are from observations with an angular resolution of 
$\sim$5$^{\prime\prime}$ compared to $\sim$1$^{\prime\prime}$ by K98. 
}
\end{figure}

\begin{table}
\caption{Positions of radio cores}
\begin{tabular}{lll}
Source     &  RA (B1950)  &  DEC (B1950)    \\
           &              &                 \\
0148$-$297 & 01 48 19.679 & $-$29 46 45.00  \\
0428$-$281 & 04 28 17.326 & $-$28 07 11.20  \\
0551$-$226 & 05 51 17.431 & $-$22 40 18.50  \\
0938$-$205 & 09 38 30.873 & $-$20 33 46.60  \\
0955$-$283 & 09 55 36.536 & $-$28 23 45.60  \\
           &              &                 \\
1022$-$250 & 10 22 57.085 & $-$25 01 08.00  \\
1023$-$226 & 10 23 10.267 & $-$22 38 07.00  \\
1026$-$202 & 10 26 35.174 & $-$20 12 04.20  \\
1029$-$233 & 10 29 12.174 & $-$23 23 56.00  \\
1107$-$218 & 11 07 44.986 & $-$21 51 13.40  \\
           &              &                 \\
1126$-$290 & 11 26 26.443 & $-$29 05 02.40  \\
1358$-$214 & 13 58 47.273 & $-$21 27 40.40  \\
\end{tabular}
\end{table}

\section{Relativistic beaming in the cores} 

To examine the consistency of the observations with expectations
of the unified scheme, one needs to estimate reliably the 
flux density of the cores at the different frequencies. 
Since our observations at all the frequencies except at
8 GHz have been made with coarser angular resolution than is
desirable, we have compared our measurements at 5 GHz with those of K98, which
were also made at about 5 GHz but with an angular resolution of 
about 1$^{\prime\prime}$. The data are plotted in Figure 1 for all the 
objects in our sample for
which core flux densities have been listed by K98.
Since the epochs of the observations 
are separated by about 5 to 10 yr, there could be differences
due to variability of the core flux density. It is, however,  clear from 
the Figure that the values are consistent over a wide range of flux density.
Besides suggesting that the cores might be only weakly variable, this suggests
that the effect of the coarser resolution is not significant at 5 GHz. 
At U band, it is likely to be even less significant, while at the L band we
have identified those which are likely to be significantly affected by diffuse
emission in the vicinity of the cores. 

\begin{figure}
\vbox{
\vspace{-1.0in}
\psfig{figure=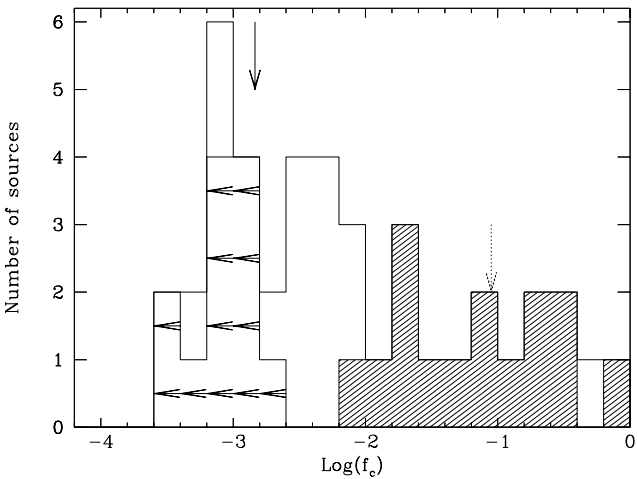,height=3.5in}
}
\caption{Distributions of f$_c$, the fraction of emission from
the core at an emitted frequency of 8 GHz for the Molonglo 
radio galaxies and quasars in our sample. The quasars are shown 
hatched, while the upper limits are indicated by horizontal 
arrows.  The median values of f$_c$ for 
radio galaxies and quasars are shown by vertical solid and 
dotted arrows respectively. 
}
\end{figure}

\subsection{Core prominence}
In the orientation-based unified scheme for radio galaxies and 
quasars, the quasars are expected to show more prominent 
cores than radio galaxies due to the effects of relativistic
beaming.  All quasars have detected cores, compared to about 50\% 
for the radio galaxies. 
The distributions of f$_c$, the fraction of emission from the core
at an emitted frequency
of 8 GHz, for our sample of radio galaxies and quasars is shown
in Figure 2. The median values of the f$_c$ are about 0.09$\pm$0.04
for quasars and less than about 0.0015$\pm$0.02 for the radio galaxies. 
Given the statistical uncertainties due to the small sample, these values are
consistent with earlier estimates for the 3CR sample (cf. Saikia \&
Kulkarni 1994) as well as for a larger sample of Molonglo objects 
(K98). These values are consistent with the unified scheme for radio galaxies 
and quasars.

\begin{figure}
\vbox{
\vspace{-0.40in}
\psfig{figure=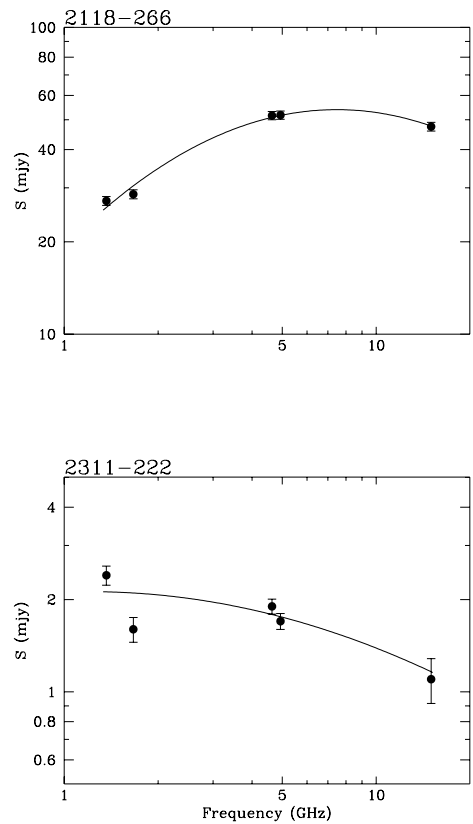,height=4.7in}
}
\caption{Observed spectra of two radio galaxy cores. The x-axis
represents the frequency in GHz, while the y-axis represents the
flux density in mJy. The source names are given in the top-left
corner of each spectrum.
The details of the fits are described in the text. }
\end{figure}

\subsection{Spectra of cores}

Since only two of the galaxies have cores detected at all
the observed wavelengths to determine reliably the spectra of
the cores (Figure 3), we concentrate on the core spectra of the
quasars to examine the effects of relativistic beaming. The
spectra of all the cores in quasars are presented in Figure 4. In
both the figures, the error bars correspond to 3\% error on
absolute flux density calibration and 1$\sigma$ noise in the image.
The spectra have been fitted by either a two- or three-degree
polynomial, and in one source, 0454$-$220, by a straight line
and a two-degree polynomial. The spectra are usually complex. 
The steep low-frequency spectra in the quasars 0058$-$229 
and 1247$-$290 are possibly due to contributions from more 
extended emission near the core. The core in   1226$-$297 is a 
candidate GPS source (cf. O'Dea 1998) with a turnover frequency
around 15 GHz, but higher frequency measurements would be
required to confirm this. 

\begin{figure*}
\vbox{
\vspace{0.30 in}
\hbox{
\hspace{-0.5in}
\psfig{figure=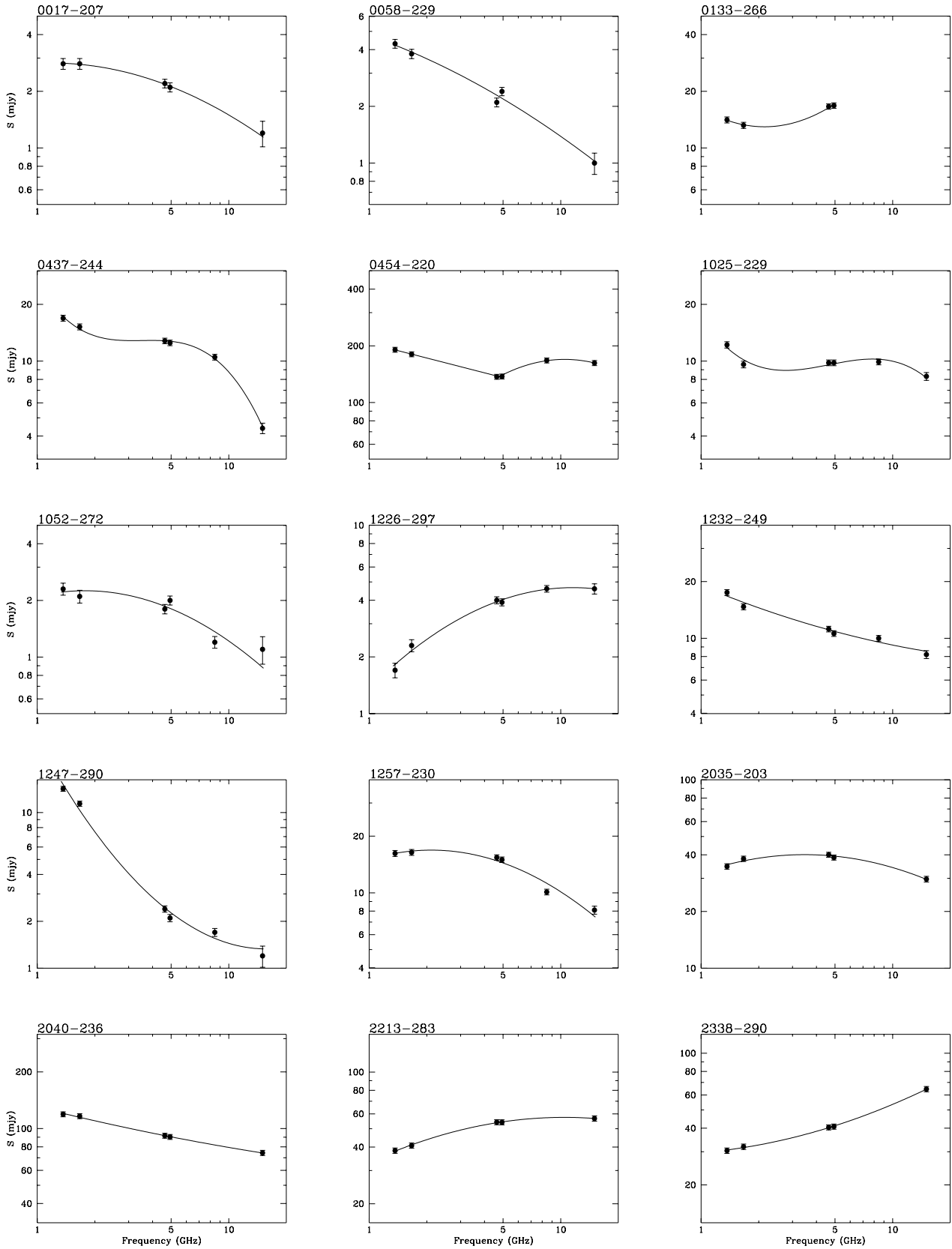,height=10.5in}
\vspace{-1.75 in}
}
}
\caption{Observed radio spectra of quasar cores. The x-axis 
represents the frequency in GHz, while the y-axis represents the
flux density in mJy. The source names are given in the top-left corner
of each spectrum.  The details of the fits are described in the text. }
\end{figure*}

\begin{figure*}
\vbox{
\vspace{-1.85 in}
\hbox{
\hspace{0.4in}
\psfig{figure=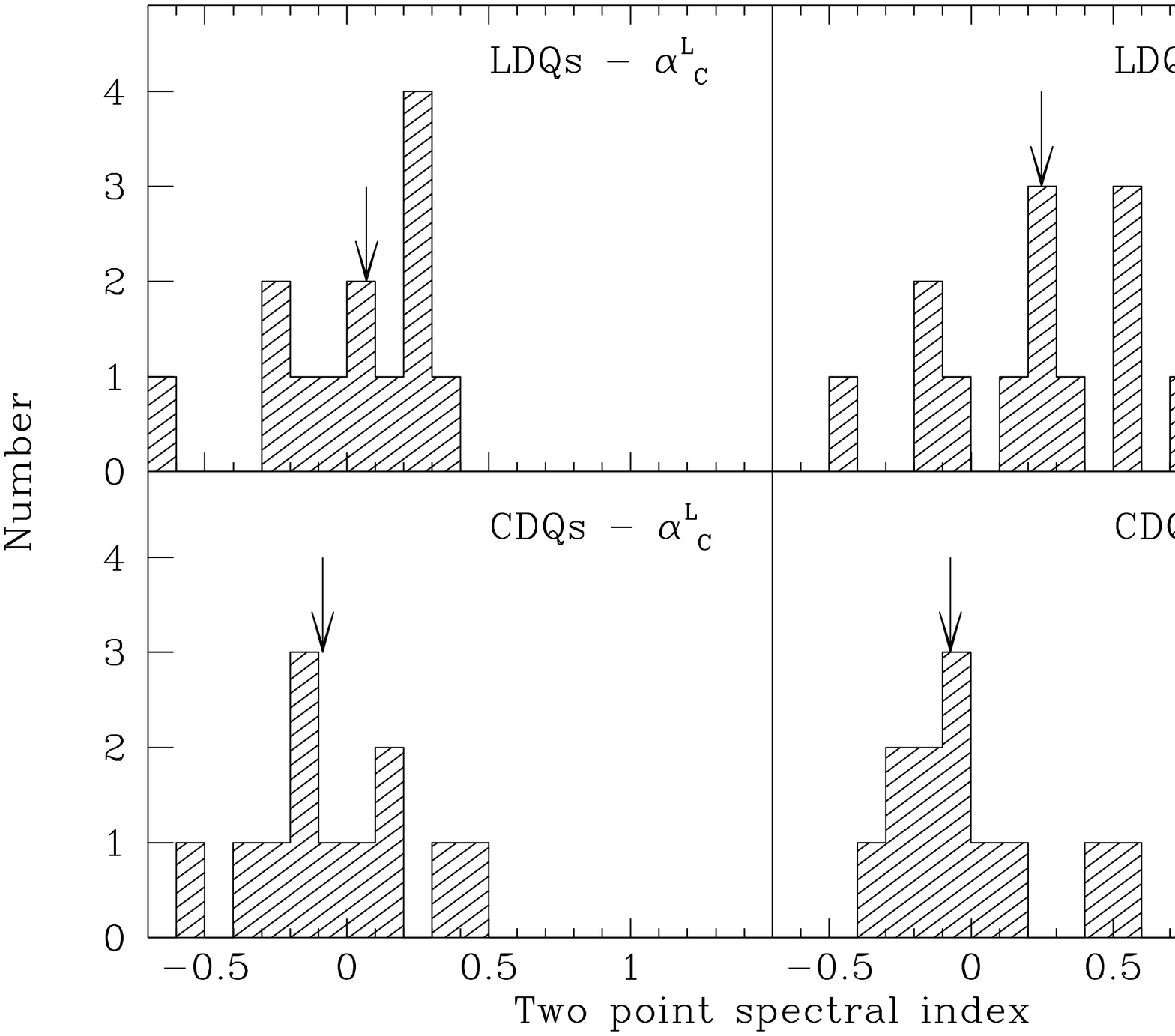,height=6.5in}
\vspace{0.1 in}
}
}
\caption{Comparison of the two-point spectral indices of 
lobe-dominated (LDQ) and core-dominated quasars (CDQ). The distributions of 
the spectral index between the L and C bands and between the C and U bands 
are shown in the left and right panels respectively.  The median 
values of each distribution are shown by arrows. 
}
\end{figure*}

In the observed spectra of the cores, the rest frequency 
in the frame of the quasar will appear shifted due to both
cosmological redshift and Doppler shift due to relativistic 
beaming of the nuclear or core emission. The rest frequency will be
shifted by an amount $\delta/(1 + z)$ where $ \delta = 
[\gamma(1 - \beta {\rm cos}\theta)]^{-1} $ is the Doppler 
factor, $z$ the redshift of the source, $\gamma$ the 
Lorentz factor, $\beta = v/c$, and $\theta$ is the angle of
inclination of the jet or source axis to the line-of-sight. 
For fixed observed frequencies, $\nu_1$ and $\nu_2$, 
the spectrum will be blue shifted for quasars, and in case 
of radio galaxies it will be generally redshifted. Assuming 
a Lorentz factor of 5, the 
Doppler factor $\delta$ for inclination angles of 
15$^\circ$, 30$^\circ$ and 60$^\circ$, which 
correspond roughly to orientations of core-dominated quasars (CDQs), 
lobe-dominated quasars (LDQs) and radio galaxies, are 3.7, 1.3 and 0.4 
respectively. Since we have determined the spectra of only 2
cores in radio galaxies, we concentrate on the quasars and
compare their spectra with a sample of core-dominated quasars
which have been observed at a single epoch by Saikia et al. (1998).
The two samples have similar redshift distributions and have 
similar extended radio luminosity within a factor of $\sim$2. 
We present the distributions of spectral index between 1.4 and 5 GHz 
($\alpha^L_C$) and that between 5 and 15 GHz ($\alpha^C_U$) for 
both the samples of LDQs and CDQs in Figure 5. These 
show that both $\alpha^L_C$ and 
$\alpha^C_U$ for CDQs are flatter compared to 
the corresponding values of LDQs, and the effect is 
more prominent in the high-frequency spectral index value
($\alpha^C_U$). The median values of the two-point spectral 
index between 1.4 and 5 GHz and that between 5 and 15 GHz for 
LDQs are 0.07 $\pm$0.074 and 0.25 $\pm$0.085 respectively, while 
the corresponding values for the CDQs are 
$-$0.085 $\pm$ 0.073 and $-$0.073 $\pm$ 0.068 respectively.
A Kolmogorov-Smirnov test shows that the distributions of 
$\alpha^C_U$ for LDQs and CDQs are different at $>$ 99\% 
significance level, while that of $\alpha^L_C$ are different 
at less than about 80\% significance level. These trends are consistent
with the unified scheme since the 
spectra of CDQs will be blue shifted by a larger amount 
than LDQs due to relativistic beaming. The trend is more 
prominent at the higher frequency, possibly due to often 
being closer to the optically thin region of the synchrotron spectrum. 

\section{Relativistic motion in the hotspots} 

Estimates of hotspot advance speeds made from attributing arm-length
asymmetries to geometrical projection and light-travel time effects
range from about 0.2$-$0.5c (e.g. Longair \& Riley 1979; Banhatti 1980;
Best et al. 1995). Tighter constraints have been placed by Scheuer (1995)
by noting that the approaching side can be identified in sources with
radio jets (cf. Saikia 1981). He estimates
an upper limit of $\sim$0.1c. The properties of highly asymmetric or completely
one-sided radio sources indicate hotspot advance speeds in the range of 
$\sim$0.2$-$0.8c
(Saikia et al. 1990). Proper motion studies of hotspots in 
compact VLBI-scale double radio sources yield values from about 
0.05-0.5c (cf.O'Dea 1998). 

\subsection{Effects of relativistic motion in hotspot spectra} 

Assuming that the hotspots are advancing at mildly
relativistic speeds, the observed spectral indices  of the oppositely directed
hotspots might appear to be different due to relativistic motion.
If the hotspot spectra steepen towards higher frequencies, the
approaching hotspot should exhibit a flatter spectrum over a fixed
observed frequency range compared to the
receding one due to relativistic Doppler effects. If the 
spectra are straight, any observed difference will reflect 
intrinsic differences in their spectra. 

\begin{figure}
\vbox{
\vspace{-1.10 in}
\psfig{figure=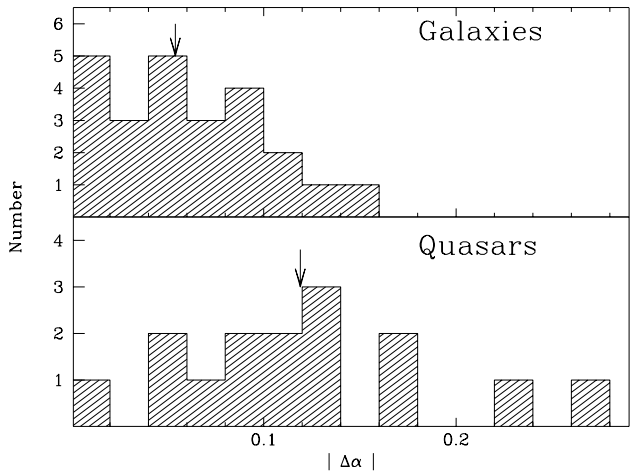,height=3.5in}
\vspace{0.1 in}
}
\caption{The distributions of the absolute difference in the hotspot spectral
index $\mid\Delta\alpha\mid$ between 1.4 and 5GHz for the radio
galaxies and quasars. The arrows indicate the median values.
}
\end{figure}

In the unified scheme for radio galaxies and quasars, the latter
should exhibit a larger difference in the spectral indices of the
hotspots on opposite sides due to smaller angles of inclination to
the line-of-sight. Assuming a canonical spectrum for the hotspot, 
one might also be able to estimate the hotspot advance speeds for
a sample of sources from the observed difference in hotspot
spectral indices. For sources with detected radio jets, one can
identify the approaching and receding hotspot. However, since 
we do not have detected radio jets in almost all
the sources in our sample, we attempt such a study by examining
the absolute difference in the hotspot spectral indices of the
oppositely directed hotspots for the radio galaxies and quasars.
In our study, we have defined the hotspot 
as the peak pixel in each lobe having surface brightness at least 10
times the weakest reliable flux density level. The number of sources
with reliable hotspot flux densities at U band are small. Therefore
we concentrate on the L- and C-band observations, but check for
consistency with the U-band observations.

Our samples of radio galaxies and quasars are of similar redshift and 
luminosity, and therefore the effects of cosmological redshift are
similar. The distributions of the absolute difference in the hotspot spectral 
index $\mid\Delta\alpha\mid$ between 1.4 and 5GHz is shown in Figure 6. 
There is a clear tendency for the
quasars to show a larger spectral index difference than radio galaxies.
The median values of $\mid\Delta\alpha\mid$ are about 0.055$\pm$0.0072
and 0.12$\pm$0.015 for the radio galaxies and quasars respectively. 
A Kolmogorov-Smirnov test shows that the distributions are different 
at $>$ 99\% significance level. The corresponding values for the 
spectral indices between C- and U-bands are 0.10$\pm$0.015 and 
0.19$\pm$0.07 for the radio galaxies and quasars, again consistent
with the trend expected in the unified scheme.

To understand the above difference in terms of relativistic beaming, 
we have considered a model hotspot spectrum with 
a curvature of 0.2 in the spectral index between each of the following
successive pairs of frequencies, namely 0.408, 1.4, 5 and 15 GHz.
There has been evidence in the past for curvature 
in the hotspot spectra. For example, Carilli et al. (1991) observed Cygnus A
with an angular resolution of about 4.$^{\prime\prime}$5 over a large frequency
range and find evidence of spectral curvature in the hotspots. Wright, Chernin
\& Forster (1997) observed the hotspots in Cygnus A with an angular resolution of
about 0.$^{\prime\prime}$4 and also find evidence of spectral steepening. (See also
Section 5.2 of this paper). The flatter spectra of the hotspots on the 
jet side can be better explained if the intrinsic spectra are curved downwards
towards higher frequencies (Dennett-Thorpe at al. 1997, 1999). 
Assuming the above spectrum and an angle of 30-50$^\circ$ for quasars 
and 50-90$^\circ$ for radio galaxies, hotspot advance speeds in the range of $\sim$0.2-0.5c 
are required to produce the observed difference in the distributions. However, with our
angular resolution of about 5$^{\prime\prime}$, the corresponding sizes are typically
about 25 kpc which is significantly larger than the sizes of hotspots in 
FRII radio sources (cf. Bridle et al. 1994; Fernini, Burns \& Perley 1997; Hardcastle et al. 1998;
Jeyakumar \& Saikia 2000). Since the spectra of the extended emission tend to be steeper
than the hotspots themselves, and the effects of relativistic beaming of hotspots 
are significant for quasars (Dennett-Thorpe at al. 1997, 1999), the relative contribution of
the extended emission to our hotspot flux density within the 5$^{\prime\prime}$ beam 
would tend to be larger for the receding hotspot compared to the approaching one.
This would tend to increase the apparent difference in the spectral indices of the
oppositely-directed hotspots, leading to a larger estimate of the hotspot velocity. 
Therefore, while our present estimates of $\sim$0.2-0.5c should be considered to be upper limits, 
a similar technique applied to higher resolution observations of hotspots should
yield more reliable estimates of hotspot speeds. From our study the hotspot speeds
appear to be at most mildly relativistic. It would also be important to 
study the curvature in the intrinsic spectra of the hotspots from higher-resolution
observations.

If the sources are intrinsically symmetric and the effects of evolution
of individual components with age are not dominant, one would expect the
approaching hotspot, which has a flatter spectral index, to be farther
from the nucleus and also brighter. However, 
we have shown in our earlier paper (IC98) that most of these sources
appear to be evolving in an asymmetric environment. Nevertheless, for
quasars where the effects of orientation are likely to be more significant,
we do find a weak trend for the flatter hotspot to be on the longer side.
This is true for 9 of the 14 quasars when one considers the spectral index
between the L and C bands, and 7 of the 8 quasars for hotspot spectral
indices between the C and U bands. The galaxies do not show any trend. 
We have also examined the relationship of the hotspot spectral indices 
with hotspot brightness ratio, but do not find any trend 
possibly due to the effects of environment as well as evolution of the 
individual components with age.

\begin{figure}
\vbox{
\vspace{-0.22in}
\psfig{figure=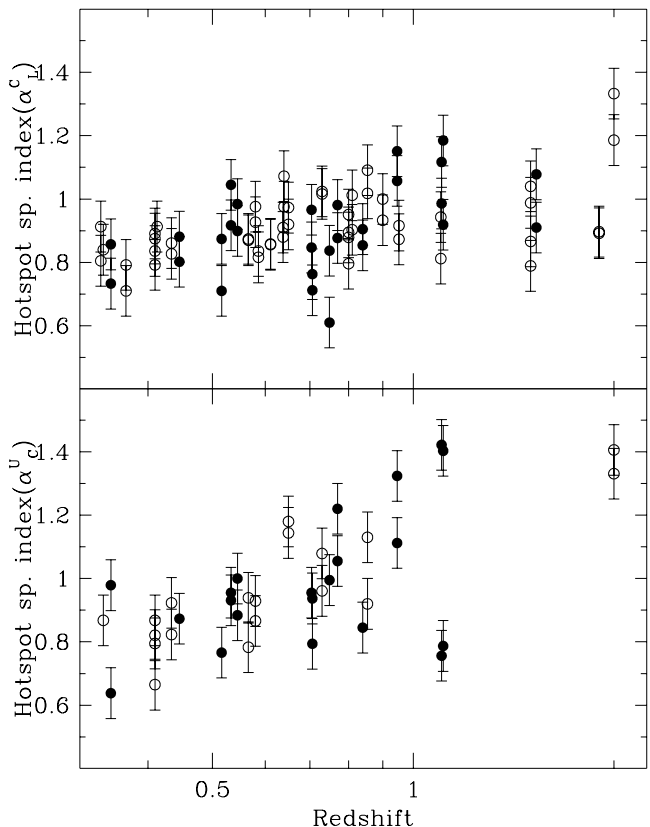,height=4.7in}
}
\caption{ Plot of hotspot spectral index against the source 
redshift. The y-axis represents the hotspot spectral index 
between 1.4 and 5GHz (upper panel) and between 5 and 15 GHz
(lower panel). Filled and open circles represent quasars and 
radio galaxies respectively. }
\end{figure}

\subsection{Spectral index - redshift/luminosity  relationship for hotspots}

We have examined the spectral index - redshift/luminosity relationship for
our sample using the spectra of the hotspots. At first glance, there appears
to be a significant
correlation of the spectral indices, $\alpha_{hs}$, between 1.4 and 5 GHz,
as well as between 5 and 15 GHz with redshift, z, the relationship being
steeper for the higher frequency spectral index (Figure 7). A Spearman
rank correlation test shows the relationship to be significant at a 
level $>$99 per cent. Since our sources are 
from a flux-density limited sample, redshift and luminosity 
are strongly correlated and we cannot distinguish between a
dependence on either luminosity or redshift.  

\begin{figure}
\vbox{
\vspace{-1.10in}
\psfig{figure=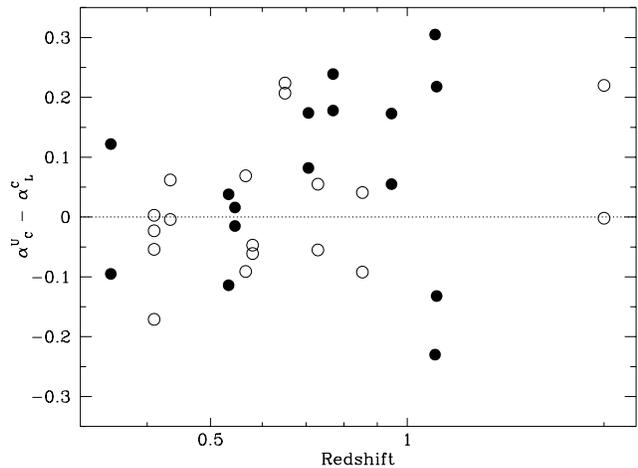,height=3.5in}
\vspace{0.2 in}
}
\caption{Plot of the difference of hotspot spectral index 
between 5 and 15GHz and that between 1.4 and 5GHz against 
the source redshift.  Filled and open circles represent quasars and
radio galaxies respectively.
}
\end{figure}

Steepening in the hotspot spectra towards higher frequencies, due to 
radiative losses can, in principle cause such a relationship since the emitted
frequency will be higher for the higher redshift and hence more luminous
objects. Inverse-Compton losses would also be important at higher redshifts. 
In a flux density limited
sample, increased magnetic field for the more luminous objects
will also increase the synchrotron emissivity, leading to an increase in
the rate at which spectral steepening occurs (cf. Laing \& Peacock 1980;
Gopal-Krishna \& Wiita 1990; Krolik \& Chen 1991; Blundell, Rawlings \& Willott 1999).
Considering all the objects with z$<$1, where the relationship is better defined, and
dividing them into two equal groups in redshift the median values of
$\alpha^U_C - \alpha^C_L$ are about 0 and 0.06. There is a weak trend for the 
high-frequency spectra to be steeper than the low-frequency ones as one goes toward
higher redshifts upto z$\sim$1 (Figure 8). Although we have plotted
the observed spectral indices rather the ones in the rest frame of the source,
the median redshift of the objects with z$<$1 is about 0.6, so that the
emitted frequencies for the low-frequency spectral index is about 2.2 and 7.8 GHz.
The median values of  $\alpha^C_L$ increase from about 0.8 to 1 as the redshift
increases from about 0.3 to 1. This range is larger than the observed degree of
spectral curvature, indicating that the correlation is not merely due to
K-correction factors alone. This is consistent with earlier suggestions
for the $\alpha$-P relationship using the integrated spectra of the sources
(cf. Lacy et al. 1993; van Breugel \& McCarthy 1989).

However, while investigating the $\alpha$-z relationship for hotspots one needs to
consider the linear resolutions of the observations which are likely to be coarser at higher
redshifts, leading to increased lobe contamination with increasing redshift. 
We have plotted the physical area of the 
restoring beam, A$_b$, for each source as a function of redshift, and have 
examined the $\alpha_{hs}$-z relationship for sources in 
restricted ranges of A$_b$ as shown in Figure 9. We first consider the spectral
indices between L and C bands where there are a larger number of sources. 
The slopes of the $\alpha_{hs}$ - z relationship for
sources with 800$<$ A$_b$ $<$1300 kpc$^2$, 1300$<$ A$_b$ $<$2000 kpc$^2$ and 
2000$<$ A$_b$ $<$2800 kpc$^2$ are 0.11$\pm$0.08, 0.21$\pm$0.11 and 0.48$\pm$0.16
respectively.  For sources observed with the highest linear resolution, which 
has the minimum contamination from the extended emission, the 
$\alpha_{hs}$ - redshift plot is relatively flat, while the slope increases for sources
observed with larger physical beam areas. A similar trend is seen while
considering the hotspot spectral indices between the C and U bands. The slopes of 
the $\alpha_{hs}$ - z plot for the three ranges of beam areas are 
0.47$\pm$0.30, 0.99$\pm$0.26 and 0.90$\pm$0.39 respectively, the errors being
larger because of the smaller number of sources observed at U band. These trends  suggest
that the hotspot spectral indices, which are closely related to the injection
spectrum, are at best weakly dependent on  redshift, while the observed relationship
is largely due to the well-known relationship for the extended emission. It would be
relevant to investigate the $\alpha_{hs}$-z relationship with higher linear
resolution than has been possible from our observations.

\begin{figure}
\vbox{
\vspace{-1.0in}
\psfig{figure=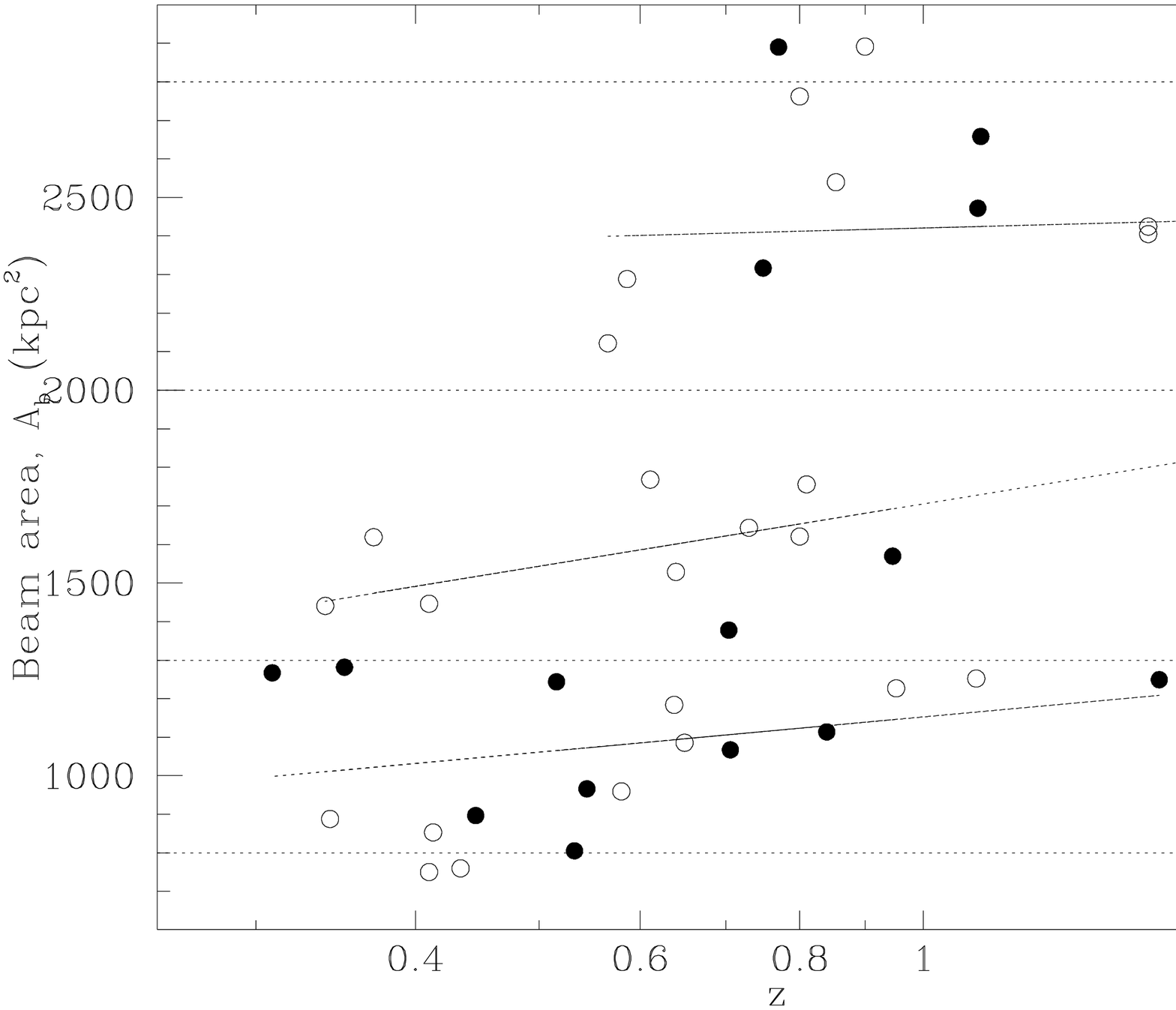,height=3.40in}
\psfig{figure=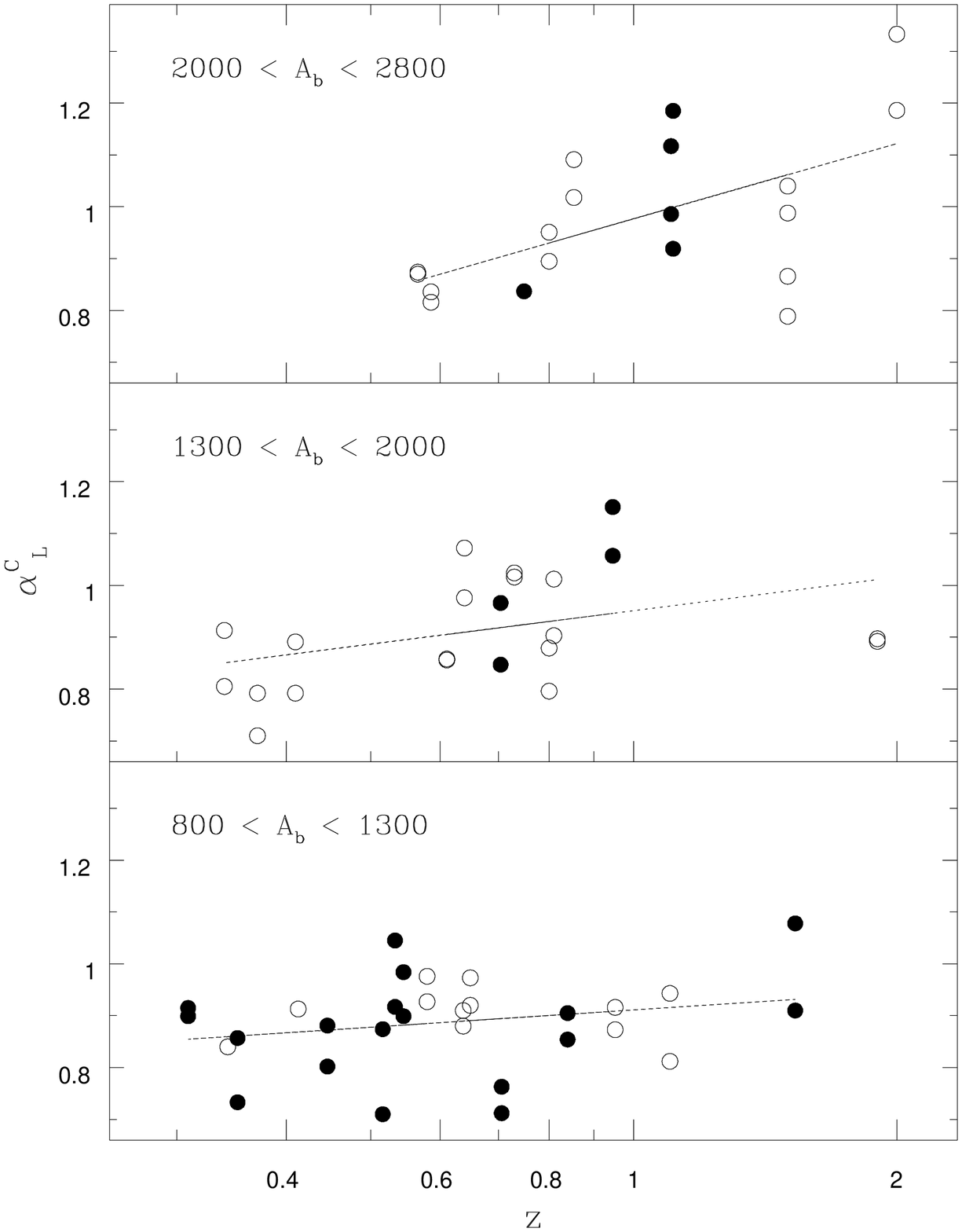,height=4.35in}
}
\caption{The area of the restoring beam, A$_b$, in units of kpc$^2$ plotted
against redshift for all the sources (upper figure). The hotspot
spectral indices between L and C bands are plotted against redshift
for sources in the three restricted ranges of beam area marked 
in the upper figure. The ranges of A$_b$ are labelled in each plot. 
The dashed lines indicate the least squares fits to the data.
Filled and open circles represent quasars and
radio galaxies respectively.
}
\end{figure}

\section{Concluding remarks}

We have studied multi-frequency radio spectra of cores and 
hotspots of a matched sample of radio galaxies and quasars. 
A comparison of the spectral indices of the cores of 
lobe-dominated and core-dominated quasars shows that the high-frequency 
spectral index of these two classes are significantly different. 
The difference can be understood in terms of Doppler effects 
and are consistent with the basic ideas of the unified scheme
for radio galaxies and quasars.

The difference in the spectral indices of the hotspots on 
opposite sides of the nucleus is larger for quasars compared
to radio galaxies, the median values of the difference being
about 0.12 and 0.06 respectively. This difference could also
be understood in terms of mild relativistic beaming of the
hotspots. The hotspots have a curved radio spectrum steepening
towards higher frequencies, possibly due to radiative losses.
The difference is consistent with the unified scheme for
radio galaxies and quasars and yields  
the velocity of advancement of the hotspots to be at most
mildly relativistic.

We have investigated the correlation between spectral index and 
redshift/luminosity for the hotspots in our sample of sources.
Although there appears to be a significant correlation, one needs
to understand the effects of spectral steepening and lobe 
contamination in the hotspot spectral indices. 
Examining the low-
and high-frequency spectral indices of the hotspots in our
sample, the observed correlation cannot be due to the effects
of K-correction alone. However, considering sources observed
with similar linear resolutions in terms of beam areas, the 
$\alpha_{hs}$-z relationship is flatter for the ones observed
with the highest linear resolutions compared to those observed
with coarser resolution. The hotspot spectral indices depend
at best marginally on redshift for those observed with the
highest linear resolutions, while for those observed with 
coarser linear resolutions the relationship appears similar to
the well-known correlation for the extended emission. It would 
be interesting to investigate the spectral index - redshift 
relationship for hotspots using higher linear resolution than has 
been possible from our observations.

\section*{Acknowledgments}
We thank N.D. Ramesh Bhat, K.S. Dwarakanath, Gopal-Krishna, Vasant
Kulkarni and Paul Wiita for their comments on the manuscript,
and Ramesh Bhat for computational help. We are also indebted 
to an anonymous referee for very helpful and critical comments
on the paper.
The National Radio Astronomy Observatory is a facility of 
the National Science Foundation operated under co-operative 
agreement by Associated Universities Inc. We thank the staff 
of the Very Large Array for the observations.

{}
\end{document}